\documentclass[RNAAS]{aastex631}

\begin{document}

\newcommand{\IAIFI}{The NSF AI Institute for Artificial Intelligence and Fundamental Interactions}

\newcommand{\MIT}{Department of Physics and Kavli Institute for Astrophysics and Space Research, Massachusetts Institute of Technology, 77 Massachusetts Avenue, Cambridge, MA 02139, USA}

\newcommand{\CfA}{Center for Astrophysics $|$ Harvard \& Smithsonian, Cambridge, MA 02138, USA}

\title{\texttt{reLAISS}: A Python Package for Flexible Similarity Searches of Supernovae and Their Host Galaxies}

\author{E.~Reynolds}
\affiliation{\CfA}

\author[0000-0003-4906-8447]{A.~Gagliano}
\affiliation{\IAIFI}
\affiliation{\CfA}
\affiliation{\MIT}

\author[0000-0002-5814-4061]{V.~A.~Villar}
\affiliation{\CfA}
\affiliation{\IAIFI}

\begin{abstract}
Discovery rates of supernovae are expected to surpass one million events annually with the Vera C. Rubin Observatory. With unprecedented sample sizes of both common and rare transient types, photometric classification alone will be insufficient for finding one-in-a-million events and prioritizing the 1\% of events for spectroscopic follow-up observations. Here, we present \texttt{reLAISS}, a modified framework for similarity searches of supernovae using extracted features of ZTF light curves and Pan-STARRS host galaxy photometry and built on the original \texttt{LAISS} framework. Unlike its predecessor, \texttt{reLAISS} couples interpretable light curve morphology features with extinction-corrected host-galaxy colors to probe both explosion physics and associated stellar populations simultaneously. The library allows users to customize the number of neighbors retrieved, the weight of host and light curve features, and the use of Monte Carlo simulations to ensure relevant matches when features are poorly constrained. We release \texttt{reLAISS} as a pip-installable package with an accompanying reference set of 20,000 features, and a set of tutorials that demonstrate the code's expanded functionality. All source code can be found at the project repository \href{https://github.com/evan-reynolds/re-laiss}{\includegraphics[height=1em]{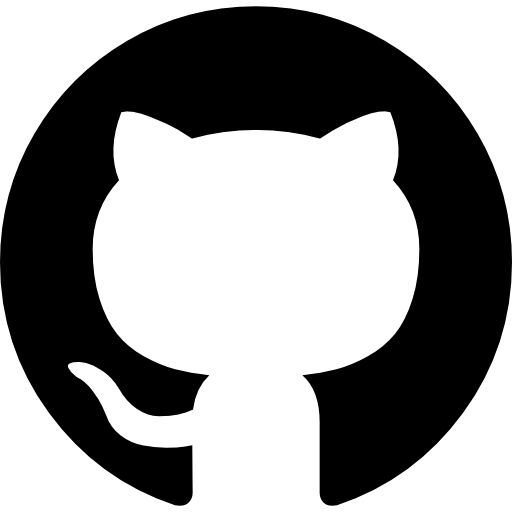}}
.
\end{abstract}


\section{Introduction} \label{sec:intro}
Wide-field surveys such as the Zwicky Transient Facility \citep{2019Bellm_ZTF} and the Young Supernova Experiment \citep{2021Jones_YSE} have cataloged the explosions of stars as supernovae (SNe) at unprecedented rates. The classification of SNe by spectroscopy and photometry has become an indispensable tool for understanding discovered events and coordinating follow-up observations. The suite of available algorithms now employ template matching \citep{2009Rodney_Templates}, feature extraction \citep{2024deSoto_Superphot}, and deep learning \citep{villar2020superraenn,2021Boone_ParSNIP} to distinguish between the explosions of old and young stars and identify energy inputs beyond $^{56}$Ni. 

As sample sizes of both common and rare types continue to swell, there is increasing demand for algorithms that can identify relevant SNe directly from the observed data instead of relying exclusively on our existing taxonomy. This may include finding in-class outliers (e.g., atypically luminous Type Ia SNe) that may signify the extremes of an evolutionary pathway. Though photometry and spectroscopy offer the most direct probe of an SN's progenitor system and underlying physics, an event's host galaxy provides broad constraints. The observational properties of a galaxy tightens the prior on the stellar population that produced a transient where event data is minimal, and transient populations obtained from host galaxy selection cuts can facilitate comparative studies of SN sub-types \citep{2025Burgaz_SNeIaHosts}.

In this work, we present \texttt{reLAISS} \citep{reLAISS_zenodo}, a library for supernova similarity searches. \texttt{reLAISS} expands upon the initial Light curve Anomaly Identification and Similarity Search (\texttt{LAISS}) library constructed by \cite{2024Aleo_reLAISS} and enables flexible and scalable searches of transients of interest. We outline the major modifications to the library from the initial publication and provide examples of its intended use cases in the following section.

\section{Methodology}
\texttt{reLAISS} matches a queried transient by the observational features of its photometry and, optionally, its host galaxy. Given a transient ZTFID, \texttt{reLAISS} first queries the ANTARES broker\footnote{\href{https://antares.noirlab.edu/}{https://antares.noirlab.edu/}} for the alert photometry in ZTF-$g$ and ZTF-$r$ filters (a user-provided photometry file can also be provided). In each filter, the code calculates the peak magnitude, time from first observation to peak, rise and decline times to half-flux, duration above half-flux, and the rolling variance. The $g-r$ color at peak, mean $g-r$ across all detections, and mean change in $g-r$ are also calculated using linear interpolation to align observations to a consistent phase. Finally, the median local curvature of photometry in each band is calculated in the 20 days prior to peak and the 20 days following peak, where the curvature is estimated by the centered second finite difference of the magnitudes:
\begin{equation}
    \kappa_i = \frac{m_{i+1} - 2m_{i} + m_{i-1}}{\left[\frac{1}{2}\left( t_{i+1} - t_{i-1}\right)\right]^2}
\end{equation}

The above features, which replace the \texttt{light-curve}-calculated features originally used in \texttt{LAISS} \citep{2024Aleo_reLAISS}, strike an optimal balance between interpretability and flexibility. 

After extracting light curve features, \texttt{reLAISS} identifies the most likely host galaxy of the transient in the Pan-STARRS 3$\pi$ catalog using the \texttt{Prost} library\footnote{\href{https://github.com/alexandergagliano/Prost}{https://github.com/alexandergagliano/Prost}}, which replaces the \texttt{GHOST} host association code used by \texttt{LAISS} \citep{2021Gagliano_GHOST}. \texttt{Prost} \citep{Gagliano2025_Prost} estimates the posterior probability of association using a Monte Carlo simulation over observed galaxy properties and user-defined priors for the true host's redshift, fractional offset, and absolute magnitude. Following association, \texttt{reLAISS} calculates the extinction-corrected Kron magnitude of the host galaxy in Pan-STARRS $griz$ and the $g-r$, $r-i$, and $i-z$ colors of the galaxy (which are independently weighted to de-prioritize outliers from catastrophic failures in color estimation). 

As with \texttt{LAISS}, \texttt{ANNOY}\footnote{\url{https://github.com/spotify/annoy}} is used to build an indexed space of reference features and conduct approximate nearest-neighbors searches via locality-sensitive hashing (details on the algorithms used for indexing and retrieval are given in \citealt{2024Aleo_reLAISS}). We provide an updated data set of 20,000 archival light curve and host galaxy features for querying. Measured uncertainties in the queried object's extracted features can be used to re-query and re-rank retrieved neighbors.

A user can choose to find neighbors by light curve properties, host properties, or both feature sets. If all properties are chosen, a weighting term can be input that allows the user to prioritize better matches along one feature set than the other. The user can define the number of neighbors to retrieve; if none are specified, the \texttt{kneed} package \citep{Satopaa_Kneed} is used to determine the inflection point in neighbor distances versus number of neighbors and a number of relevant neighbors is suggested. 

\texttt{reLAISS} can be used to find similar events to an observed transient or the closest matches to a theoretical model of interest. A user can also manually specify an observed host galaxy, in cases where a theoretical model is queried and correlations with a particular stellar population are desired. We now fit a Generalized Pareto distribution to the upper-tail of the nearest-neighbor distance distribution of the reference data bank; the real-time feature probabilities associated a queried transient are converted with a sigmoid transformation to an anomaly score between 0 and 100. This replaces the previous random forest algorithm within LAISS for real-time anomaly detection. Future work will be devoted to investigating the robustness of the chosen feature set and anomaly detection method for partial-phase and low-cadence light curves, and extending the features to other photometric passbands and surveys beyond ZTF. 

\begin{figure}[ht]
    \centering
    \includegraphics[width=\linewidth]{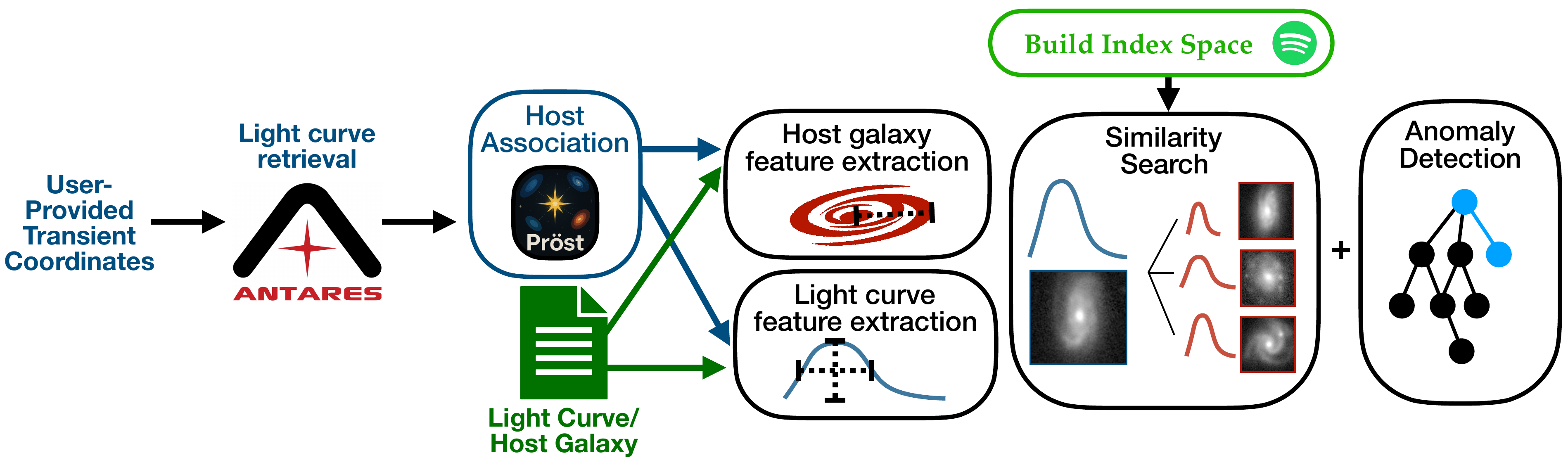}
    \vspace{1em}
    \begin{minipage}[c]{0.59\linewidth}
        \centering
        \includegraphics[width=\linewidth]{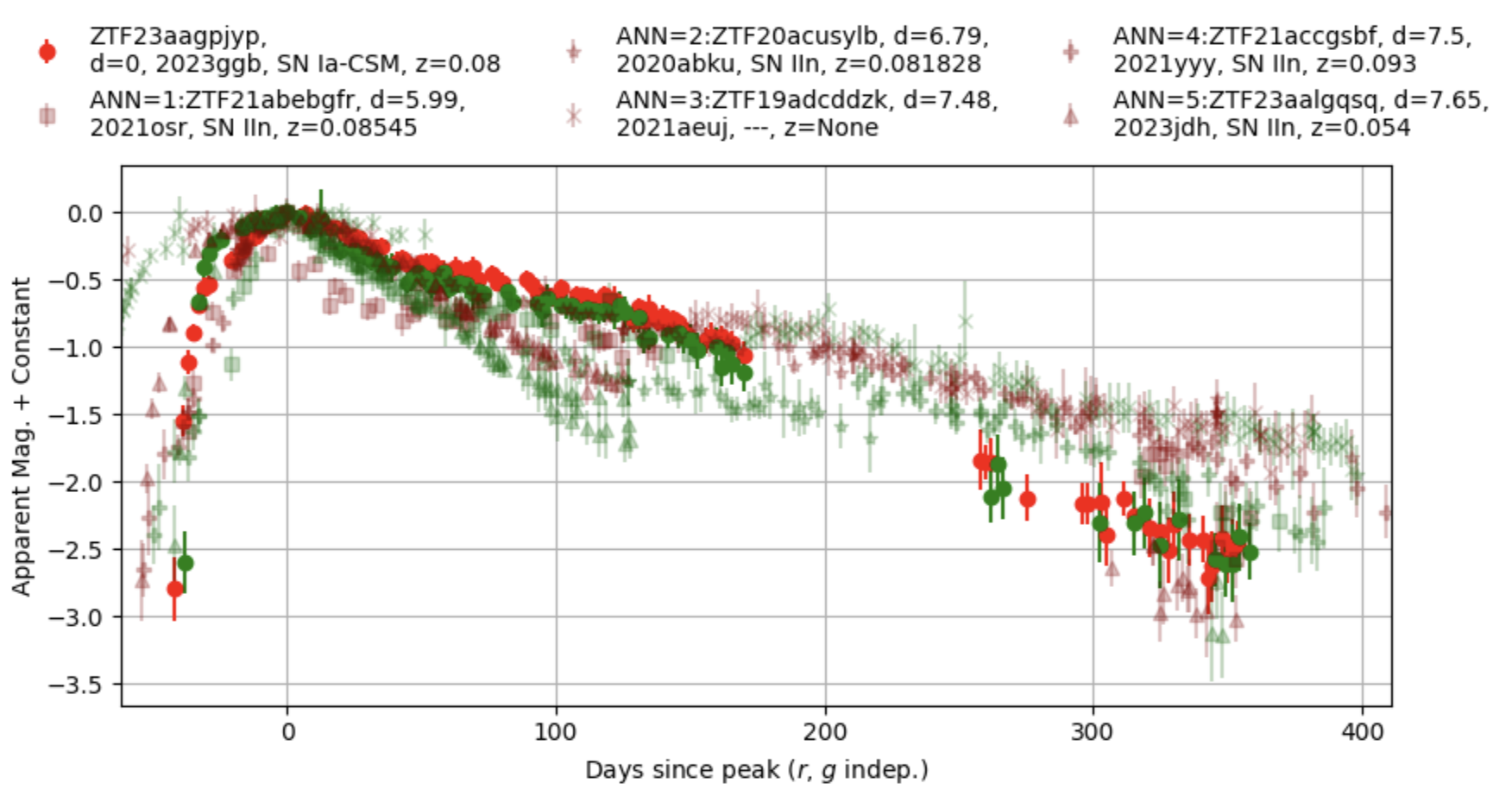}
        \textbf{(a)} Light curve neighbors
    \end{minipage}
    \begin{minipage}[c]{0.39\linewidth}
        \centering
        \includegraphics[width=\linewidth]{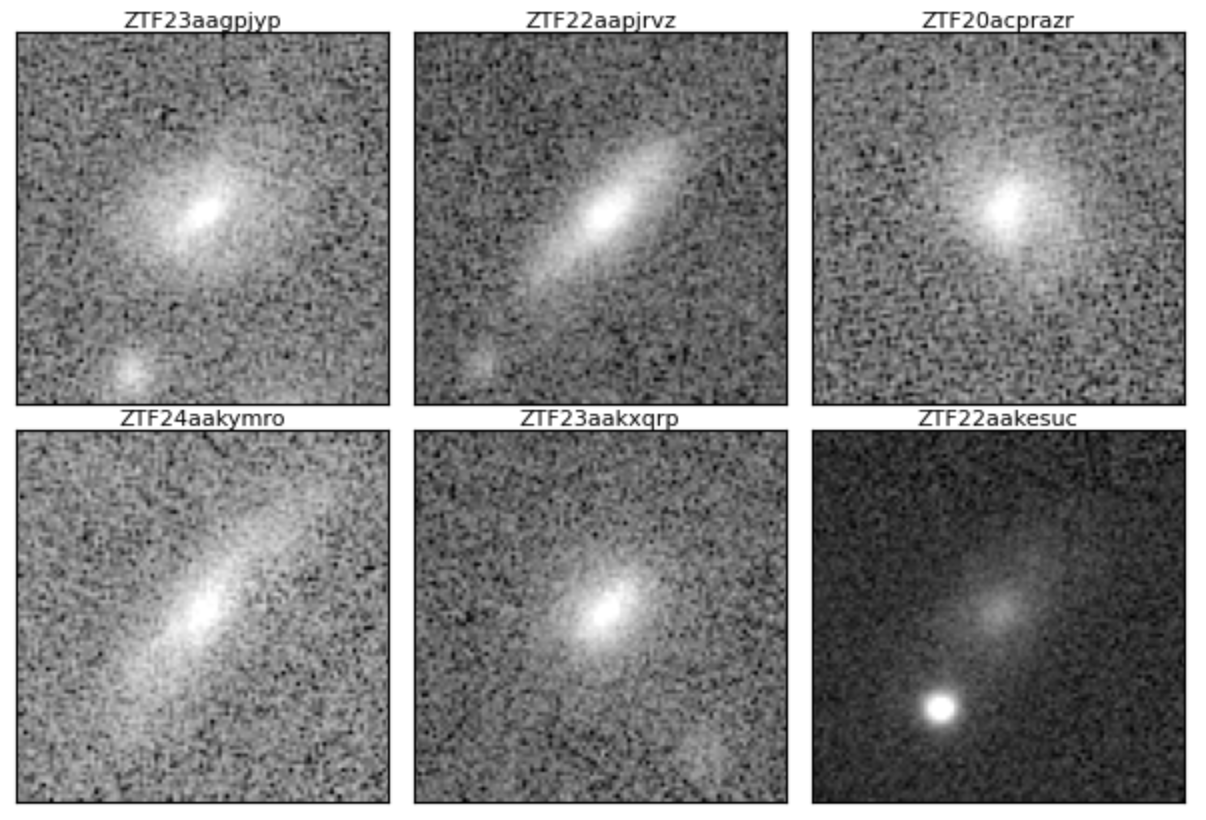}
        \textbf{(b)} Host galaxy neighbors
    \end{minipage}
    \vspace{2em}
    \begin{minipage}[c]{\linewidth}
        \centering
        \includegraphics[width=\linewidth]{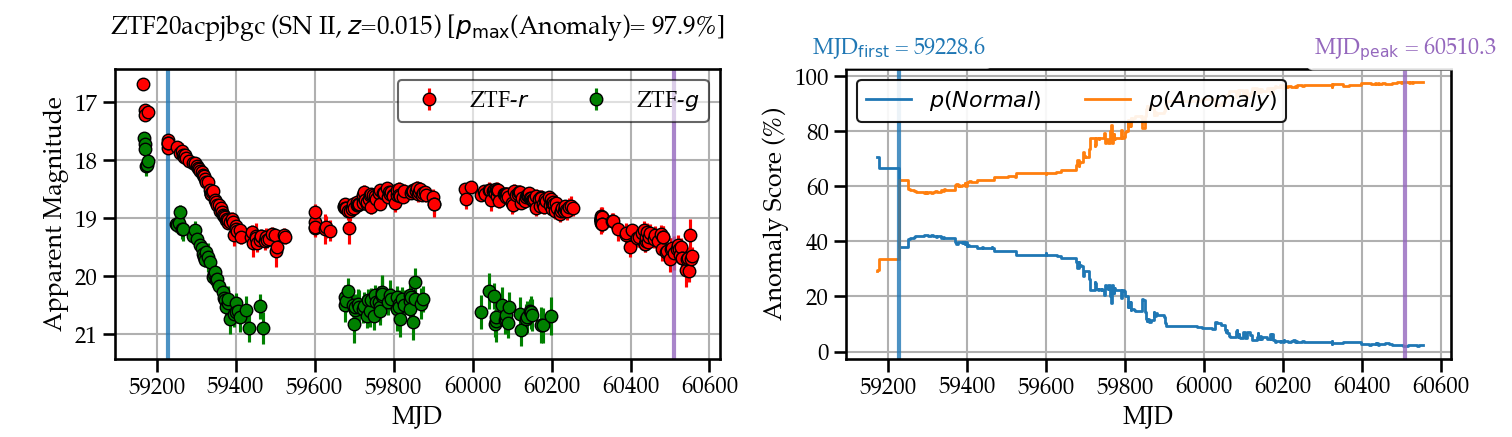}
        \textbf{(c)} Light curve for anomaly detection (left) and calibrated anomaly score (right)
    \end{minipage}

    \caption{\textbf{Top:} Workflow of the \texttt{reLAISS} library. Starting from user-provided transient coordinates or an optional light curve file, the pipeline (i) identifies the most probable Pan-STARRS host using \texttt{Prost}, (ii) extracts a suite of interpretable light curve and host-galaxy features, (iii) indexes and queries these features with locality‑sensitive hashing via \texttt{ANNOY}, and (iv) returns nearest neighbors and highlights anomalies for follow‑up. \textbf{Middle:} Example nearest neighbor searches. Panel (a) overlays light curves of neighbors for ZTF23aagpjyp, an SN~Ia-CSM, identified using light curve features only. Panel (b) charts neighbor host galaxies of the same transient identified using host galaxy features only. Each search used 20 monte carlo simulations without PCA. \textbf{Bottom:} Example anomaly detection for anomalous transient. Panels (c) and (d) show the light curve of transient ZTF20acpjbgc (a long-duration SN~II) and its time-evolving anomaly score.}
    \label{fig:relaiss_workflow}
\end{figure}

\section{Acknowledgements}
This work is supported by the National Science Foundation under Cooperative Agreement PHY-2019786 (The NSF AI Institute for Artificial Intelligence and Fundamental Interactions, http://iaifi.org/).

\bibliography{sample631}{}
\bibliographystyle{aasjournal}

\end{document}